\documentclass{article}
\usepackage{spconf,amsmath,graphicx}
\usepackage{graphicx}
\usepackage{subfigure} 
\usepackage{amsfonts}
\usepackage{booktabs}
\usepackage{multicol}
\usepackage{multirow}

\usepackage{amsmath,amssymb,amsfonts}

\begin{document}

\title{DRVC: A Framework of Any-to-Any Voice Conversion with Self-Supervised Learning}

\name{Qiqi Wang$^{1,2}$, Xulong Zhang$^{1}$, Jianzong Wang$^{1*}$, Ning Cheng$^{1}$, Jing Xiao$^{1}$
\thanks{$^*$ Corresponding author: Jianzong Wang, jzwang@188.com. This paper is supported by the Key Research and Development Program of Guangdong Province No. 2021B0101400003 and the National Key Research and Development Program of China under grant No. 2018YFB0204403.}}
\address{$^{1} $Ping An Technology (Shenzhen) Co., Ltd., China \\ $^{2} $University of Auckland, New Zealand}
%
%
%
%

%
\maketitle
\begin{abstract}
Any-to-any voice conversion problem aims to convert voices for source and target speakers, which are out of the training data. Previous works wildly utilize the disentangle-based models. The disentangle-based model assumes the speech consists of content and speaker style information and aims to untangle them to change the style information for conversion. Previous works focus on reducing the dimension of speech to get the content information. But the size is hard to determine to lead to the untangle overlapping problem. We propose the Disentangled Representation Voice Conversion (DRVC) model to address the issue. DRVC model is an end-to-end self-supervised model consisting of the content encoder, timbre encoder, and generator. Instead of the previous work for reducing speech size to get content, we propose a cycle for restricting the disentanglement by the Cycle Reconstruct Loss and Same Loss. The experiments show there is an improvement for converted speech on quality and voice similarity. 
\end{abstract}
\section{Introduction}
\label{sec:intro}


Voice conversion (VC) aims to generate a new voice with the source voice content and target speaker timbre \cite{DBLP:journals/taslp/LiuCWWLM21,tang2022avqvc,DBLP:conf/icassp/ChenSH21,DBLP:conf/icassp/HayashiHKT21}. VC models can be roughly named as $multiple_1$-to-$multiple_2$ models, with $multiple_1, multiple_2 \in \{one, many, any\}$, the $multiple_1, multiple_2$ represents the source speakers and the target speakers, respectively. $One$ means the speaker is fixed, whether the training or inferring process. $Many$ and $any$ represents the speaker is seen or unseen in the training process, respectively.

\textit{One-to-one} VC model is inefficient due to only being able to convert voice between a fixed pair of source speaker and target speaker, such as the CycleGAN-VC  \cite{asru2021zhang,DBLP:journals/taslp/NakashikaTA15,asru2021tang}. Even though the \textit{any-to-one} and  \textit{many-to-one} can work in uncertain source speaker\cite{DBLP:conf/icmcs/SunLWKM16}, but the target speaker is also fixed. For VC models with uncertain speaker pair, such as \textit{many-to-many}~\cite{DBLP:journals/access/LeeKP21,DBLP:conf/icassp/WangY21}  and \textit{any-to-any} ~\cite{DBLP:conf/icassp/LinCLLL21}, widely utilize the disentanglement-based method. Disentanglement-based models assume that the speech consists of the content and speaker style information. They aim to split the two pieces of information from speeches and exchange the content to achieve the conversion task. But the challenge is to avoid the overlapping of untangling results~\cite{DBLP:conf/iclr/YuanCZHGC21,DBLP:conf/icassp/LiTYWXSM21}. AutoVC proposes to circumspection choose the content dimension to separate the content information before combining it with the pre-trained speaker information~\cite{DBLP:conf/icml/QianZCYH19}. But it is hard to determine the number of reduced sizes to avoid residual the source speaker information or loss of the content. The similar problem also exists in VQVC+, which proposes to use a codebook to obtain the content information by combining similar dimensions~\cite{DBLP:conf/interspeech/WuCL20}. The suitable codebook size is the key factor to get mostly content information without speakers' influence.


The image-to-image (I2I) task aims to convert the target image style to the source image. Disentangled Representation for Image-to-Image Translation (DRIT) assumes image consists of content and attribute information, and two input images have same content \cite{DBLP:journals/ijcv/LeeTMHLSY20}. Besides, it novelty utilizes double exchange process for changing the content information, one for synthesis new image, one for reconstructing image, to reduce the overlapping of disentanglement.

Inspired by the double exchange process of DRIT, we propose to use the process to address the untangle overlapping problem without circumspection choose the content size. The proposed end-to-end framework is Disentangled Representation Voice Conversion (DRVC). Comparing to DRIT, we believe neither of the content or style information is the same between the input speeches. Furthermore,  we design a cycle framework for the double exchange of the style information with cycle loss and two discriminators. We experiment with the model on VCC2018, both the subjective and objective results show our model has better performance.

\section{Proposed Model}

\begin{figure}
    \centering
    \includegraphics[width=0.45\textwidth]{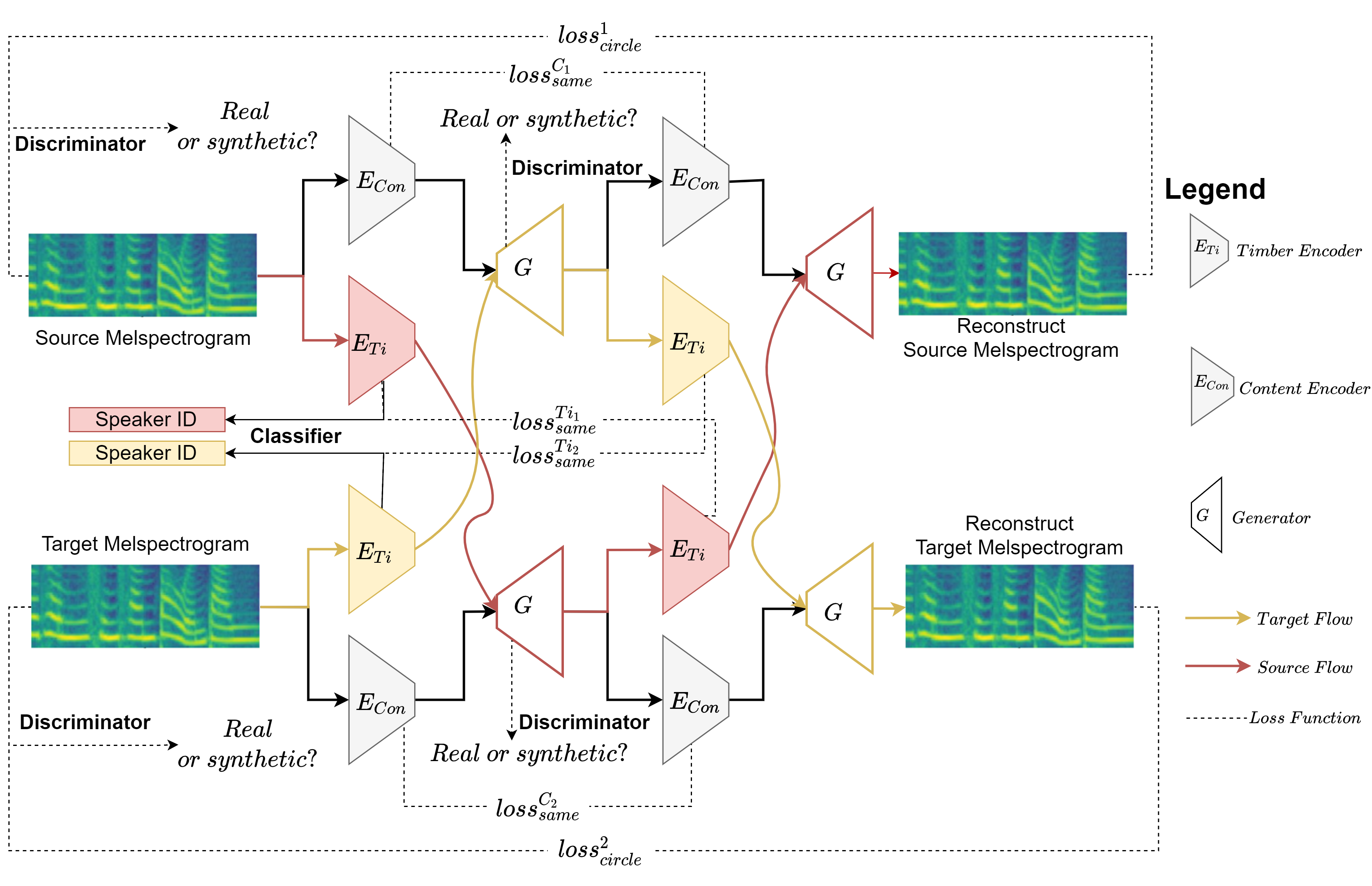}
    \caption{Architecture of the Disentangled Representation Voice Conversion (DRVC) model.}
    \label{fig:training}
\end{figure}

In this section, we will introduce the proposed Disentangled Representation Voice Conversion (DRVC). The proposed architecture for voice conversion is shown in Fig.\ref{fig:training}. 

\subsection{Overall Architecture}

The proposed DRVC model consists the content encoder $E_{Con}$, speaker style encoders $E_{S}$, generators $G$, voice discriminator $D_{v}$, and domain classifier $D_{S}$. Take the target mel-spectrogram $B$ as an example, the content encoder $E_{Con}$ map the melspectrogram into content representation ($E_{Con}: B \rightarrow C_{B}$) and the timbre encoder $E_{S}$ map the mel-spectrogram into timbre representation ($E_{S}: B \rightarrow S_{B}$). The content encoder and generator structure are the same, which consists of three CNN layers with the head and the tail by LSTM layer. We utilize the AdaIN-VC model speaker encoder \cite{DBLP:conf/interspeech/ChouL19} as the style encoder. The voice discriminator $D_{v}$ aims to distinguish the input voice is real or synthesis voice. The domain classifier $D_{S}$ aims to identify the embedding speaker style information belongs which speaker. The voice discriminator and domain classifier are multilayer perceptron with two hidden layer. As the voice conversion target, the generator $G$ synthesize the voice conditioned on both content and timbre vectors ($G: {[C_{A},S_{B}]} \rightarrow \hat{B}$).



\subsection{Disentangle Content and Style Representations}
We assume two input voices, $a$ and $b$, are spoken by two speakers, $A$ and $B$. We define the speaker $A$ is source speaker, which provide the content for converted speech. And the style information of converted speech is given by speaker $B$, who is the target speaker.

The proposed DRVC model embeds input voice melspectrograms onto specific content spaces, $C_{A},C_{B}$, and specific style spaces, $S_{A},S_{B}$. It means the content encoder embeds the content information from input speeches, and the timbre encoders should map the voices to the specific style information.
\begin{equation}
    \begin{split}
        \{\boldsymbol{a_C}, \boldsymbol{a_{S}}\} = \{E_{Con}({a}),E_{S}({a})\}, \ \boldsymbol{a_C} \in C_{A}, \boldsymbol{a_{S}} \in S_{A} \\
        \{\boldsymbol{b_C}, \boldsymbol{b_{S}}\} =  \{E_{Con}({b}),E_{S}({b})\}, \ \boldsymbol{b_C} \in C_{B}, \boldsymbol{b_{S}} \in S_{B}
    \end{split}
    \label{Eq:1}
\end{equation}
\noindent where, $a$ and $b$ represent the input source voice mel-spectrogram and target voice mel-spectrogram, respectively.

We apply two strategies to achieve representation disentanglement and avoid the overlapping problem: same embedding losses and a domain discriminator. The content and speaker style information should be unaltered regardless of the embedding process and input speeches.
\begin{equation}
    \begin{split}
        \mathcal{L}^{C_n}_{same} = \mathbb{E}[|\boldsymbol{a_C} - \boldsymbol{\tilde{a_C}}|] , \
        \mathcal{L}^{S_n}_{same} = \mathbb{E}[|\boldsymbol{a_S} - \boldsymbol{\tilde{a_S}}|]
    \end{split}
    \label{Eq:2}
\end{equation}
\noindent where, $\mathcal{L}^{C_n}_{same}$, $\mathcal{L}^{S_n}_{same}$ are the same loss of content and style information, $n \in \{a,b\} $ represents the source and target domains, respectively. And, $\tilde{a_C}$ and $\tilde{a_S}$ means the content and style information after second conversion, respectively. For sum of content same loss, $\mathcal{L}^{C}_{same} = \sum^2_n\mathcal{L}^{C_n}_{same}$, and sum of style same loss, $\mathcal{L}^{S}_{same} = \sum^2_n\mathcal{L}^{S_n}_{same}$. The total same loss is $\mathcal{L}_{same} = \mathcal{L}^{C}_{same} + \mathcal{L}^{S}_{same}$. 

The domain classifier $D_{v}$ aims to identify input style hidden vector belongs to which speaker.
\begin{equation}
    p_a = D_{v}(\boldsymbol{a_S}),\ p_b = D_{v}(\boldsymbol{b_S})
\end{equation}
\begin{equation}
    \small
    \mathcal{L}_{domain} = -\frac{1}{2}(\sum_{i} y_{a}(i)log(p_{a}(i)) + \sum_{i} y_{b}(i)log(p_{b}(i)))
\end{equation}
\noindent  where, $y$ is the real target, $p$ is the predicted target.


\subsection{Cycle Loss}
The proposed model performs voice conversion by combining the source voice content $\boldsymbol{a_C}$ and the target voice timbre $\boldsymbol{b_{S}}$. Similar to the CycleGAN-VC, we propose to use a cycle process (i.e., $B \rightarrow \tilde{A} \rightarrow \hat{B}$) to train the generator $G$. And we will exchange the timbre and content information twice. Furthermore, we use a cross-cycle consistency as a loss function $\mathcal{L}_{cycle}$. 

\textbf{First conversion.} Given a non-corresponding pair of voices' mel-spectrogram $a$ and $b$, we have the content information $\{ \boldsymbol{a_{C}, b_{C}} \}$, and style information $\{ \boldsymbol{a_{S}, b_{S}} \}$. Then we exchange the style information $\{ \boldsymbol{a_{S}, b_{S}} \}$ to generate the $\{ \tilde{a}, \tilde{b} \}$, where $\tilde{a} \in Target \ domain, \tilde{b} \in Source \ domain$.
\begin{equation}
    \begin{split}
        \tilde{b} = G(\boldsymbol{b_{C}}, \boldsymbol{a_{S}}) \\
        \tilde{a} = G(\boldsymbol{a_{C}}, \boldsymbol{b_{S}}) 
    \end{split}
\end{equation}

\textbf{Second conversion.} To encode the $\tilde{b}$ and $\tilde{a}$ into $\{\boldsymbol{\tilde{b}_{C}, \tilde{b}_{S}}\}$ and $\{\boldsymbol{\tilde{a}_{C}, \tilde{a}_{S}}\}$, we swap the style information $\{\boldsymbol{\tilde{a}_{S}, \tilde{b}_{S}}\}$ again to converse the generated voices from first instance.
\begin{equation}
    \begin{split}
        \hat{a} = G(\boldsymbol{\tilde{a}_{C}}, \boldsymbol{\tilde{b}_{S}}) \\
        \hat{b} = G(\boldsymbol{\tilde{b}_{C}}, \boldsymbol{\tilde{a}_{S}}) 
    \end{split}
\end{equation}

After the two stages conversion, the output $\hat{a}$ and $\hat{b}$ should be the reconstruct of the input $a$ and $b$. In other words, the best target of the relation between the input and output is $\hat{a} = a$ and $\hat{b} = b$. We use the cross-cycle consistency loss $\mathcal{L}_{cross}$ to enforce this constraint.
\begin{equation}
    \begin{split}
        \mathcal{L}_{cycle} = \mathbb{E}_{a,b}[||G(E_{Con}(\tilde{a}), E_{S}(\tilde{b})) - a||_{1} \\ + ||G(E_{Con}(\tilde{b}), E_{S}(\tilde{a})) - b||_{1}]
    \end{split}
\end{equation}

Furthermore, to avoid the generator over-fitting, we propose an identity loss. Identity loss aims to restrict the generator synthesize original speech when inputting the original speech content and style information.
\begin{equation}
    \begin{split}
        \mathcal{L}_{id} = \mathbb{E}_{a,b}[||G(E_{Con}(a), E_{S}(a)) - a||_{1} \\ + ||G(E_{Con}(b), E_{S}(b)) - b||_{1}]
    \end{split}
\end{equation}

\subsection{Adversarial Loss}
Inspired by the GAN, we utilize an adversarial loss $\mathcal{L}_{adv}$ to enforce the generated speech to be sound like natural speech.

We train the voice discriminator by directly input the real voice $\{a,b\}$ and synthesis speeches $\{\tilde{a},\tilde{b}\}$. We set the synthesis speech is fake, and natural speech is real. Besides, we add a Gradient Reversal Layer in the discriminator.
\begin{equation}
    F(\tfrac{\partial \mathcal{L}_{c}}{\partial \theta_{G}}) = -\lambda ( \tfrac{\partial \mathcal{L}_{R}}{\partial \theta_{G}} + \tfrac{\partial \mathcal{L}_{F}}{\partial \theta_{G}})
\end{equation}
\noindent where, $F(\cdot)$ is the mapping function of gradient reversal layer,
$\lambda$ is the weight adjustment parameters, $\theta_{G}$ is the parameter of the generator. And, $\mathcal{L}_{R}$, $\mathcal{L}_{F}$ are the classification loss of real and fake, respectively. $R$ represents real, and $F$ means fake.

The adversarial loss $\mathcal{L}_{adv}$ is,
\begin{equation}
    \begin{split}
        \mathcal{L}_{adv} =  \mathbb{E}_{R\sim p(a)}[logD_{S}(a)]  + \mathbb{E}_{F\sim p(\tilde{a})}[logD_{S}(\tilde{a})] \\ +  \mathbb{E}_{R\sim p(b)}[logD_{S}(b)]  + \mathbb{E}_{F\sim p(\tilde{b})}[logD_{S}(\tilde{b})]
    \end{split}
\end{equation}
\noindent where, real A speech $a$, real B speech $b$, fake A speech $\tilde{a}$, and fake B speech $\tilde{b}$ are trained the discriminator.

The full objective function of the DRVC is:
\begin{equation}
    \begin{split}
        \mathcal{L}_{all} =  \lambda_{cycle} \mathcal{L}_{cycle} + \lambda_{id} \mathcal{L}_{id} + \lambda_{S} \mathcal{L}_{adv}  \\+ \lambda_{domain} \mathcal{L}_{domain} + \lambda_{same} \mathcal{L}_{same}
    \end{split}
\end{equation}
\noindent where, the $\mathcal{L}_{all}$ is the total loss of this framework. The $\lambda_{cycle}$, $\lambda_{id}$, $\lambda_{S}$, $\lambda_{domain}$ and $\lambda_{same}$ represent the weight of each loss.

\section{Experiments}
\label{sec:Ex}

\subsection{Dataset}
We conduct experiments on the VCC2018 dataset \cite{DBLP:conf/odyssey/Lorenzo-TruebaY18}, which professional US English speakers record. There are four females and four males voices as sources, and two females and two males voices as targets. Each speaker speeches are divided into 35 sentences for evaluation and 81 sentences for training. All speech data is sampling at 22050 Hz. We utilize all speakers except VCC2SF3, VCC2TF1, VCC2SM3, and VCC2TM1 speakers to train the DRVC model. The remain speakers is used to test the model performance on any-to-any phase. Besides, we choose VCC2SF4, VCC2TF2, VCC2SM4, and VCC2TM2 speakers to test for many-to-many phase.

\subsection{Model Configuration}
Our proposed model was trained on a single NVIDIA V100 GPU. We set that the $\lambda_{cycle} = 5 $, $\lambda_{id} = 2$, $\lambda_{S} = 1$, $\lambda_{domain}=10$ and $\lambda_{same}= 50$. Meanwhile, the decay of the learning rate is pointed at $5*10^{-6}$ every epoch. Following [35], we gradually changed the parameter $\lambda = = \tfrac{2}{1+exp(-10*k)} - 1$ in speaker classifier, where $k$ is the percentage of the training process. We utilize the Adam optimizer with $\beta_1 = 0.9$, $\beta_2 = 0.99$, $\epsilon = 10^{-9}$. Besides, we utilize the Mel-GAN as the vocoder.

We utilize the offical codes of AutoVC~\cite{DBLP:conf/icml/QianZCYH19}, VQVC+~\cite{DBLP:conf/interspeech/WuCL20}, and AGAIN-VC~\cite{DBLP:conf/icassp/ChenWWL21} as the baselines. We strict follow the instruction of the provided codes by the authors and use the same training and testing database with us.

\subsection{Subjective Evaluation Setting}

We set two experiments. The first one aims to evaluate the model any-to-any performance. There are four subtests, Female to Female (VCC2SF3-VCC2TF1), Female to male (VCC2SF3-VCC2TM1), Male to Female (VCC2SM3-VCC2TF1), and Male to Male (VCC2SM3-VCC2TM1). The second aims to test the model on many-to-many phase. The test includes  Female to Female (VCC2SF4-VCC2TF2), Female to male (VCC2SF4-VCC2TM2), Male to Female (VCC2SM4-VCC2TF2), and Male to Male (VCC2SM4-VCC2TM2). We evaluate the synthetic voice performance by using the Mel-cepstral distortion (MCD) \cite{DBLP:conf/interspeech/BrandtZAM17}. Besides, we also set a subject evaluation tests for Voice Similarity (VSS) and the speech quality on Mean Opinion Score (MOS).

Both MOS and VSS are obtained by asking 30 people with an equal number of gender to rate the output audio clips. About the knowledge background,  testers have different knowledge fields, such as  Computer Vision, Human Resources, Psychology, etc. Listeners can give zero to five marks to show how they feel the voice is clear (five means the best) on the MOS. On the VSS test, listeners need to fill the blank to choose one of the most similar synthesis voices to real or choose none of them is similar.

\begin{table}[]

  \centering
  \fontsize{8}{7}\selectfont
  \setlength{\abovecaptionskip}{0pt}%
  \setlength{\belowcaptionskip}{10pt}%
  \caption{Comparison of different models in any-to-any and many-to-many. $\Downarrow$ means lower score is better, and $\Uparrow$ means bigger score is better.}
  \label{Comparison}
    \begin{tabular}{ccccc}
    \toprule
    \multirow{2}{*}{\textbf{Methods}}&
    \multicolumn{2}{c}{\textbf{Any-to-Any}}&\multicolumn{2}{c}{\textbf{ Many-to-Many}}\cr
    \cmidrule(lr){2-3} \cmidrule(lr){4-5}
    & MCD $\Downarrow$ & MOS$\Uparrow$ & MCD$\Downarrow$ & MOS$\Uparrow$ \cr
    \midrule
    Real & - & 4.65 $\pm$ 0.12 
    & - 
    & 4.66 $\pm$ 0.21\cr
    VQVC+ & 7.47 $\pm$ 0.07 & 2.52 $\pm$ 0.42 
    & 7.78 $\pm$ 0.07 
    & 2.62 $\pm$ 0.22\cr
    AutoVC & 7.69 $\pm$ 0.21 & 2.95 $\pm$ 0.56
    & 7.61 $\pm$ 0.17 
    & 3.17 $\pm$ 0.65\cr
    AGAIN-VC& 7.42 $\pm$ 0.19 & 2.45 $\pm$ 0.34
   & 7.64 $\pm$ 0.21 & 2.47 $\pm$ 0.58\cr
    \midrule
    \textbf{DRVC} &\textbf{7.39 $\pm$ 0.05} &\textbf{3.32 $\pm$ 0.36} &\textbf{7.59 $\pm$ 0.04} 
    &\textbf{3.51 $\pm$ 0.52}\cr
    \bottomrule
    \end{tabular}
    \label{ta:1}
\end{table}

\begin{figure}
    \centering
    \subfigure[Any-to-any phase]{\label{fig:subfig:a}
\includegraphics[width=0.98\linewidth]{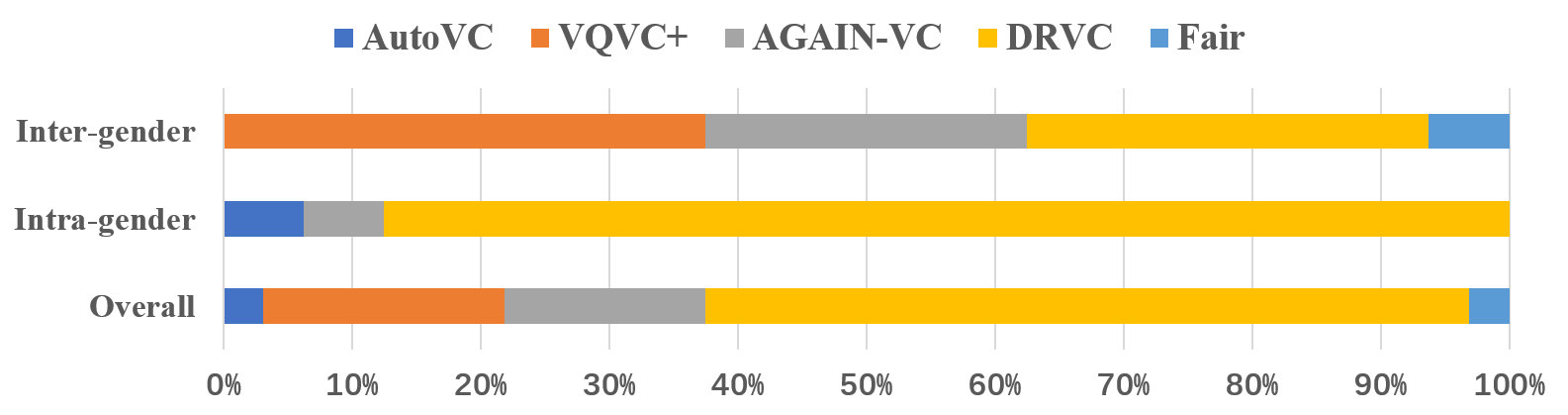}}
\vfill
\subfigure[Many-to-many phase]{\label{fig:subfig:b}
\includegraphics[width=0.98\linewidth]{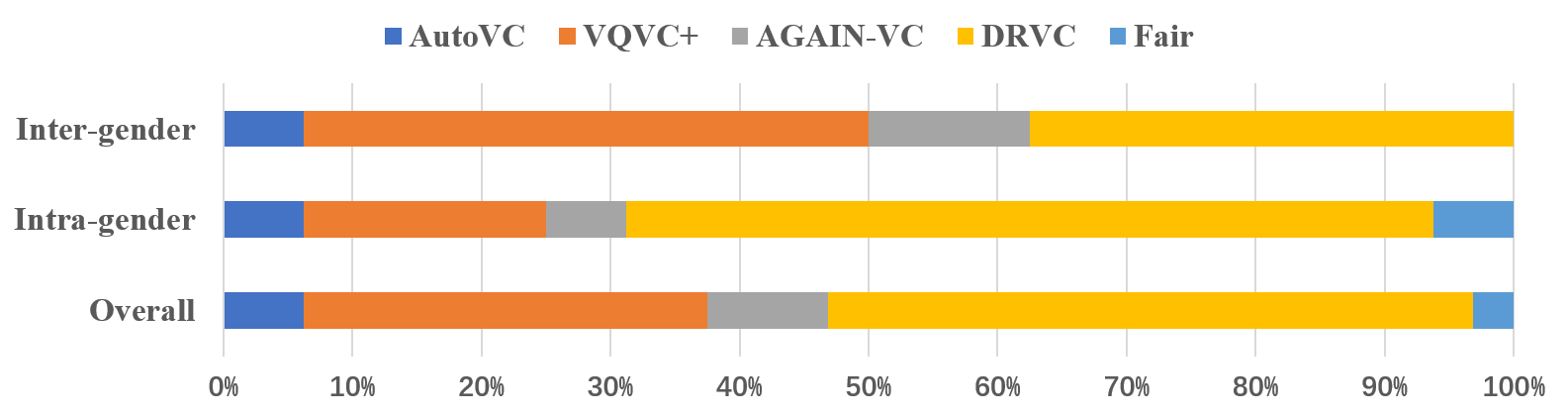}}
    \caption{Comparison of voice similarity on different models}
    \label{fig:VSS}
\end{figure}


\subsection{Result and Discussion}
\label{sec:RD}
Table \ref{ta:1} shows different models' MCD and MOS results on both any-to-any and many-to-many phases. Table \ref{tab:2} shows the ablation experiments result for the proposed model. Figure \ref{fig:VSS} shows different models' voice similarity result.

\textbf{Overall result} Both subjective and objective shows the proposed model achieves better performance. Our model average improves by about 0.4 marks in MOS and 0.05 marks on MCD. In inter-gender voice conversion, the performance of DRVC is similar to the AGAIN-VC. Also, the MCD results prove AGAIN-VC and MCD have close performance. But the proposed model also gets a little better improvement on intra-gender voice conversion. Figure \ref{fig:VSS} shows the proposed model has much better performance. Because most of the listeners believe the synthesis speeches made by DRVC are similar to the target speeches. Table \ref{ta:1} shows the performance of the baselines is lower than the previously reported. The previous works are trained based on the VCTK database, including 109 speakers and hundreds of utterances~\cite{VCTK}. But the VCC2018 only consists of eight speakers and 116 utterances per speaker is much smaller than the VCTK. In other words, the utilized database is a low-resource situation. It is why the baselines are low-performance than their reports.

\textbf{Any-to-Any} Figure \ref{fig:VSS} and Table \ref{ta:1} show the proposed method has outhperformance on all the three evaluation metrics. Especially, in the intra-gender experiments, most of the listeners believe the synthesis speeches by the proposed model are closly to the orginal speeches.

\textbf{Many-to-Many} Figure \ref{fig:VSS} shows all of these models have a better performance on the many-to-many phase. Because the target speaker is already seen in the training process, it will be easy to synthesize speech. However, the many-to-many MCD result is better than any-to-any. Due to the MCD calculates the differences on two mel-spectrograms and mel-spectrograms can not represent the voice naturalness. It may have different evaluation results. The MOS test shows the many-to-many has better performance. Besides, the proposed model has a bigger ratio on any-to-any in VSS evaluation. But it not represents the proposed model worse in the many-to-many phase. Due to all methods are improved, respondents may have disagreements. The number of respondents who chose baseline is increasing. Even though the disagreement exists, the proposed model also has a little better performance on the many-to-many phase.

\textbf{Ablation experiments} We set the ablation experiments to compare the MCD results by removing the used loss functions in DRVC. Table \ref{tab:2} shows to utilize the full loss functions is much better than delete any one of them. Besides, as expected to remove the Domain Loss or Voice Same Loss pose the highest MCD result. Because the group of Domain Loss and Voice Same Loss is used to restrict the speeches with the same speaker have the same disentanglement results. The interesting finding is when removing the adversarial loss also leads to the high MCD problem. The reason is the synthesis speech without the adversarial loss restriction is unnatural.  These speeches may have more different frames than natural speeches. The sounds of synthesis audio prove this reason which is the most unnatural of these experiments. 

\begin{table}[]
    \centering
    \small
    \caption{Ablation experiments on the proposed model. $\Downarrow$ means lower score is better.}
    \begin{tabular}{cc}
    \toprule
        Model & MCD$\Downarrow$ \\
        \hline
        DRVC \textit{w/o} Cycle Loss & 7.68 $\pm$ 0.26 \\
        DRVC \textit{w/o} Identity Loss & 7.63 $\pm$ 0.14 \\
        DRVC \textit{w/o} Domain Loss & 7.72 $\pm$ 0.12 \\
        DRVC \textit{w/o} Voice Same Loss & 7.75 $\pm$ 0.32 \\
        DRVC \textit{w/o} Content Same Loss & 7.50 $\pm$ 0.32 \\
        DRVC \textit{w/o} Adversarial Loss & 7.72 $\pm$ 0.35 \\
        \hline
        \textbf{DRVC} & \textbf{7.39 $\pm$ 0.05} \\
        \toprule
    \end{tabular}
    \label{tab:2}
\end{table}
\section{Conclusion}
\label{sec:Co}


We propose the Disentanglement Representative Voice Conversion (DRVC) framework to address the disentanglement overlapping problem and avoid subjectively choosing the content size. DRVC uses a cycle process, and a series of untangling loss functions to restrict the content and style information is non-overlapping. The experiment results with the VCC2018 dataset demonstrate that DRVC better performance on MOS and MCD.
\section{Acknowledgement}

This paper is supported by the Key Research and Development Program of Guangdong Province No. 2021B0101400003 and the National Key Research and Development Program of China under grant No. 2018YFB0204403.





\clearpage
\bibliographystyle{IEEEbib}
\bibliography{refs}

\begin{thebibliography}{10}

\bibitem{DBLP:journals/taslp/LiuCWWLM21}
Songxiang Liu, Yuewen Cao, Disong Wang, Xixin Wu, Xunying Liu, and Helen Meng,
\newblock ``Any-to-many voice conversion with location-relative
  sequence-to-sequence modeling,''
\newblock {\em {IEEE} {ACM} Trans. Audio Speech Lang. Process.}, vol. 29, pp.
  1717--1728, 2021.

\bibitem{tang2022avqvc}
Huaizhen Tang, Xulong Zhang, Jianzong Wang, Ning Cheng, and Jing Xiao,
\newblock ``Avqvc: One-shot voice conversion by vector quantization with
  applying contrastive learning,''
\newblock in {\em 2022 IEEE International Conference on Acoustics, Speech and
  Signal Processing (ICASSP2022)}. IEEE, 2022, pp. 1--5.

\bibitem{DBLP:conf/icassp/ChenSH21}
Mingjie Chen, Yanpei Shi, and Thomas Hain,
\newblock ``Towards low-resource stargan voice conversion using weight adaptive
  instance normalization,''
\newblock in {\em {ICASSP}}, 2021, pp. 5949--5953.

\bibitem{DBLP:conf/icassp/HayashiHKT21}
Tomoki Hayashi, Wen{-}Chin Huang, Kazuhiro Kobayashi, and Tomoki Toda,
\newblock ``Non-autoregressive sequence-to-sequence voice conversion,''
\newblock in {\em {ICASSP}}, 2021, pp. 7068--7072.

\bibitem{asru2021zhang}
Xulong Zhang, Jianzong Wang, Ning Cheng, Edward Xiao, and Jing Xiao,
\newblock ``{CycleGEAN}:cycle generative enhanced adversarial network for voice
  conversion,''
\newblock in {\em {IEEE} Automatic Speech Recognition and Understanding
  Workshop (ASRU2021)}. 2021, pp. 1--6, {IEEE}.

\bibitem{DBLP:journals/taslp/NakashikaTA15}
Toru Nakashika, Tetsuya Takiguchi, and Yasuo Ariki,
\newblock ``Voice conversion using {RNN} pre-trained by recurrent temporal
  restricted boltzmann machines,''
\newblock {\em {IEEE} {ACM} Trans. Audio Speech Lang. Process.}, vol. 23, no.
  3, pp. 580--587, 2015.

\bibitem{asru2021tang}
Huaizhen Tang, Xulong Zhang, Jianzong Wang, Ning Cheng, Zhen Zeng, Edward Xiao,
  and Jing Xiao,
\newblock ``{TGAVC}: Improving autoencoder voice conversion with text-guided
  and adversarial training,''
\newblock in {\em {IEEE} Automatic Speech Recognition and Understanding
  Workshop (ASRU2021)}. 2021, pp. 1--6, {IEEE}.

\bibitem{DBLP:conf/icmcs/SunLWKM16}
Lifa Sun, Kun Li, Hao Wang, Shiyin Kang, and Helen~M. Meng,
\newblock ``Phonetic posteriorgrams for many-to-one voice conversion without
  parallel data training,''
\newblock in {\em {ICME}}, 2016, pp. 1--6.

\bibitem{DBLP:journals/access/LeeKP21}
Yun~Kyung Lee, Hyun~Woo Kim, and Jeon~Gue Park,
\newblock ``Many-to-many unsupervised speech conversion from nonparallel
  corpora,''
\newblock {\em {IEEE} Access}, vol. 9, pp. 27278--27286, 2021.

\bibitem{DBLP:conf/icassp/WangY21}
Chao Wang and Yibiao Yu,
\newblock ``Non-parallel many-to-many voice conversion using local linguistic
  tokens,''
\newblock in {\em {ICASSP}}, 2021, pp. 5929--5933.

\bibitem{DBLP:conf/icassp/LinCLLL21}
Yist~Y. Lin, Chung{-}Ming Chien, Jheng{-}Hao Lin, Hung{-}yi Lee, and Lin{-}Shan
  Lee,
\newblock ``Fragmentvc: Any-to-any voice conversion by end-to-end extracting
  and fusing fine-grained voice fragments with attention,''
\newblock in {\em {ICASSP}}, 2021, pp. 5939--5943.

\bibitem{DBLP:conf/iclr/YuanCZHGC21}
Siyang Yuan, Pengyu Cheng, Ruiyi Zhang, Weituo Hao, Zhe Gan, and Lawrence
  Carin,
\newblock ``Improving zero-shot voice style transfer via disentangled
  representation learning,''
\newblock in {\em ICLR}, 2021.

\bibitem{DBLP:conf/icassp/LiTYWXSM21}
Zhonghao Li, Benlai Tang, Xiang Yin, Yuan Wan, Ling Xu, Chen Shen, and Zejun
  Ma,
\newblock ``Ppg-based singing voice conversion with adversarial representation
  learning,''
\newblock in {\em {ICASSP}}, 2021, pp. 7073--7077.

\bibitem{DBLP:conf/icml/QianZCYH19}
Kaizhi Qian, Yang Zhang, Shiyu Chang, Xuesong Yang, and Mark
  Hasegawa{-}Johnson,
\newblock ``Autovc: Zero-shot voice style transfer with only autoencoder
  loss,''
\newblock in {\em {ICML}}, 2019, vol.~97 of {\em Proceedings of Machine
  Learning Research}, pp. 5210--5219.

\bibitem{DBLP:conf/interspeech/WuCL20}
Da{-}Yi Wu, Yen{-}Hao Chen, and Hung{-}yi Lee,
\newblock ``{VQVC+:} one-shot voice conversion by vector quantization and u-net
  architecture,''
\newblock in {\em Interspeech}, 2020, pp. 4691--4695.

\bibitem{DBLP:journals/ijcv/LeeTMHLSY20}
Hsin{-}Ying Lee, Hung{-}Yu Tseng, Qi~Mao, Jia{-}Bin Huang, Yu{-}Ding Lu,
  Maneesh Singh, and Ming{-}Hsuan Yang,
\newblock ``{DRIT++:} diverse image-to-image translation via disentangled
  representations,''
\newblock {\em Int. J. Comput. Vis.}, vol. 128, no. 10, pp. 2402--2417, 2020.

\bibitem{DBLP:conf/interspeech/ChouL19}
Ju{-}Chieh Chou and Hung{-}yi Lee,
\newblock ``One-shot voice conversion by separating speaker and content
  representations with instance normalization,''
\newblock in {\em Interspeech}, 2019, pp. 664--668.

\bibitem{DBLP:conf/odyssey/Lorenzo-TruebaY18}
Jaime Lorenzo{-}Trueba, Junichi Yamagishi, Tomoki Toda, Daisuke Saito, Fernando
  Villavicencio, Tomi Kinnunen, and Zhen{-}Hua Ling,
\newblock ``The voice conversion challenge 2018: Promoting development of
  parallel and nonparallel methods,''
\newblock in {\em Odyssey}, 2018, pp. 195--202.

\bibitem{DBLP:conf/icassp/ChenWWL21}
Yen{-}Hao Chen, Da{-}Yi Wu, Tsung{-}Han Wu, and Hung{-}yi Lee,
\newblock ``Again-vc: {A} one-shot voice conversion using activation guidance
  and adaptive instance normalization,''
\newblock in {\em Proc. {ICASSP}}, 2021, pp. 5954--5958.

\bibitem{DBLP:conf/interspeech/BrandtZAM17}
Erika Brandt, Frank Zimmerer, Bistra Andreeva, and Bernd M{\"{o}}bius,
\newblock ``Mel-cepstral distortion of german vowels in different information
  density contexts,''
\newblock in {\em {Interspeech}}, 2017, pp. 2993--2997.

\bibitem{VCTK}
Christophe Veaux, Junichi Yamagishi, Kirsten MacDonald, et~al.,
\newblock ``Superseded-cstr vctk corpus: English multi-speaker corpus for cstr
  voice cloning toolkit,''
\newblock 2016.

\end{thebibliography}

\end{document}